\documentclass[sn-mathphys,Numbered]{sn-jnl}

\usepackage{graphicx}%
\usepackage{multirow}%
\usepackage{amsmath,amssymb,amsfonts}%
\usepackage{amsthm}%
\usepackage{mathrsfs}%
\usepackage[title]{appendix}%
\usepackage{xcolor}%
\usepackage{textcomp}%
\usepackage{manyfoot}%
\usepackage{booktabs}%
\usepackage{algorithm}%
\usepackage{algorithmicx}%
\usepackage{algpseudocode}%
\usepackage{listings}%
\def\vc#1{\mbox{\boldmath $#1$}}
\usepackage{color}
\def\tred#1{\textcolor{black}{#1}}

\theoremstyle{thmstyleone}%
\theoremstyle{thmstyletwo}%

\theoremstyle{thmstylethree}%

\begin{document}

\title[Article Title]{Variational calculations of symmetric nuclear matter and pure neutron matter with the tensor-optimized Fermi Sphere (TOFS) method\\
{\large $-$ \tred{many-body effects and short-range correlation} $-$}
}

\author[1]{\fnm{Taiichi} \sur{Yamada}}\email{yamada@kanto-gakuin.ac.jp}



\affil[1]{\orgdiv{Faculty of Science and Engineering}, \orgname{Kanto Gakuin University}, \orgaddress{ \city{Yokohama}, \postcode{236-8501}, \country{Japan}}}




\abstract{
The equations of state for symmetric nuclear matter and pure neutron matter are investigated with the tensor-optimized Fermi Sphere method (TOFS) up to the density $\rho=0.5$~fm$^{-3}$.
This method is based on a linked-cluster expansion theorem, and the energy per particle of nuclear matter ($E/A$) is calculated variationally with respect to the correlated nuclear matter wave function.
We can study the density dependence of the many-body terms arising from the operator products, which contribute to $E/A$.
\tred{In order to clarify the relation between the many-body effects and short-range correlation, we take the spin-isospin dependent central {\it NN} interaction with a few GeV repulsion in the inner region.} 
The EOS obtained by the TOFS method is reasonably reproduced, compared with other \textit{ab initio} many-body methods.
We found that the many-body terms \tred{(from the 2-body to 6-body ones)} give sizable effects on $E/A$ at higher density, and they play an important role in nuclear matter.
}




\maketitle

\section{Introduction}\label{sec1}



It is one of the challenging problems in nuclear physics to understand the properties of nuclear matter from the first principle calculations.
The equation of state (EOS) for nuclear matter at high density is closely related to the structure of neutron star~\cite{lattimer00}, and at lower density it connects to the $\alpha$-particle condensation in symmetric nuclear matter~\cite{roepke98}, together with the $\alpha$ cluster states in finite nuclear system~\cite{schuck16}. 
The typical \textit{ab initio} methods for nuclear matter calculations are the Brueckner-Hartree-Fock (BHF)~\cite{bombaci05,song98,baldo12}, the Brueckner-Bethe-Goldstone (BBG)~\cite{song98,baldo12,baldo01}, the Fermi hypernetted chain/single-operator chain (FHNC/SOC) ~\cite{pandharipande79,fantoni98,lovato11,akmal98}, the self-consistent Green's function (SCGF)~\cite{dickhoff04,soma06,rios09}, the auxiliary field diffusion Monte Carlo (AFDMC)~\cite{schmidt99,sarsa03,gandolfi14}, the coupled-cluster (CC)~\cite{baardsen13,hagen14,lietz17}, \tred{the lowest-order constrained variational (LOCV)}~\cite{bordbar97,bordbar98,modarres23}, and so on.
The comparative study of the EOS's obtained by BHF, BBG, FHNC/SOC, SCGF, and AFDMC, using the Argonne two-nucleon potentials (AV4', AV6', AV8' and AV18)~\cite{wiringa95}, has disclosed~\cite{baldo12} that the EOS for symmetric nuclear matter depends significantly on the \textit{ab initio} methods, in particular, at the higher density region.
On the other hand, for pure neutron matter, the dependence on the methods is relatively small for any Argonne potential, although the discrepancies among them are gradually enhanced at higher density.
Recently, benchmark calculations (BBG, FHNC/SOC, AFDMC) for pure neutron matter have been performed~\cite{piarulli20,lovato22} with the chiral effective field theory potentials, as well as the AV18 force plus three-nucleon force.
The dependence of the EOS's on the calculated methods is likely to come from the treatment of the non-central components (in particular, tensor force) in medium, as well as the many-body effects appearing in nuclear matter.
Therefore, it is quite important to develop new \textit{ab initio} methods and study the EOS for nuclear matter by them.

The tensor-optimized Fermi Sphere (TOFS) method~\cite{yamada19_cluster,yamada19,yamada21} has been proposed recently, which is an \textit{ab initio} approach for nuclear matter calculations.
In this method, an exponential-type correlated nuclear matter wave function, $\exp(F) \Phi_{0}$, is taken, where the correlation function $F$ is arbitrary, for example, can induce central, tensor, etc.\ correlations, and $\Phi_{0}$ is the uncorrelated Fermi-gas wave function. 
Since the TOFS framework, which is \textit{`Hermitian'}, is guaranteed by a linked-cluster expansion theorem~\cite{yamada19_cluster}, only the linked diagrams can be taken into account in the calculation of the energy per nucleon for nuclear matter ($E/A$).

One of the particular features of TOFS is that $E/A$ is evaluated by an energy variation with respect to the correlated nuclear matter wave function.
The FHNC/SOC method, adopting the Jastow-type correlated nuclear matter wave function, is also variational.
In this approach, however, the variational calculation is performed for only the two-body cluster term in the FHNC framework, and the many-body effects in nuclear matter are estimated by the hypernetted chain summation techniques, which collects only the SOC (single-operator chain) approximable diagrams of the \textit{`nodals'} ones~\cite{pandharipande79,fantoni98,lovato11,akmal98}.
In other word, this approach is not fully variational with respect to the correlated nuclear matter wave function.

Another interesting feature of TOFS is that one can estimate the \tred{respective contributions} of each many-body term in nuclear matter \tred{with no restriction of the particular diagrams such as \textit{'nodals'}.}
\tred{The many-body terms in TOFS} are originated from the operator product, $\exp(F^{\dagger}){\mathcal H}\exp(F)$, where ${\mathcal H}$ denotes the Hamiltonian of nuclear matter.
\tred{In the FHNC/SOC framework, it is difficult to calculate the respective contributions from each manu-body term, although only the total sum of the contributions from the many-body terms more than 3-body term is estimated by the SOC equation evaluating only the SOC approximable diagrams~\cite{pandharipande79,fantoni98,lovato11,akmal98}.}
\tred{Since the respective contributions of each many-body term are largely affected by the effect of the short-range correlation and tensor correlation, they are useful quantities to see the appearance of the nucleon correlation in nuclear matter.
However, the density dependence of the respective contributions and its relation with the nucleon correlations have not been so clear even for the case of the short-range correlation.} 


\tred{
The purposes of this article are twofold.
The first is to study the EOS's of symmetric nuclear matter and pure neutron matter with TOFS up to the density $\rho=0.5$~fm$^{-3}$, using the AV4' potential, which is the spin-isospin dependent central {\it NN} interaction with a couple of GeV repulsion in the inner region. 
We will compare the results of TOFS with those of the other \textit{ab initio} methods.
The second purpose is to study the density dependence of the respective contributions from each many-body term (2-body, 3-body, 4-body, 5-body, and 6-body ones) in symmetric nuclear matter and pure neutron matter, focusing on their relation with the short-range correlation.
Since the AV4' force is central, the effects of the many-body terms seem to be induced dominantly by the short-range correlation.
Their information is useful to understand the relationship between the short-range correlation and the many-body terms, in particular, at higher density region.}
Although the partial results for only symmetric nuclear matter up to $\rho=0.2$~fm$^{-3}$ are shown in Refs.~\cite{yamada19_cluster,yamada19,yamada21}, the TOFS framework is \tred{developed} to enable to treat higher density region ($\rho \simeq 3\rho_0$, $\rho_0$ is normal density) and also pure neutron matter.

The present paper is organized as follows:\ 
In Sec.~\ref{sec2}, we formulate the TOFS approach. 
\tred{The results of EOS's obtained by TOFS are presented in Sec.~\ref{sec3}} for symmetric nuclear matter and pure neutron matter, compared with the other methods (BHF, BBG, SCGF, AFDMC, and FHNC/SOC).
\tred{We discuss the density dependence of the respective contributions from each many-body term.} 
Finally the summary is given in Sec.~\ref{sec4}.

\section{Formulation}\label{sec2}

In the TOFS method, we take the exponential type of the correlated symmetric nuclear matter wave function,
\begin{eqnarray}
&&\tred{\Psi}_{\rm ex} = \exp(F) \Phi_0, 
\label{eq:correlation_fun}\\
&&F = F_S + F_D.
\label{eq:F}
\end{eqnarray}
The uncorrelated wave function of symmetric nuclear matter $\Phi_0$ is described by the Fermi gas model, where $A$ nucleons occupy up to the Fermi sea with the Fermi wave number $k_F$,
\begin{eqnarray}
\Phi_{0} = \frac{1}{\sqrt{A!}} \det \left| \phi_{\gamma_1}(1) \phi_{\gamma_2}(2) \cdots \phi_{\gamma_A}(A) \right|.
\end{eqnarray}
The single-nucleon wave function $\phi_\gamma$ is written as
\begin{eqnarray}
&&\phi_{\gamma_n}(n) = \phi_{\vc{k}_n}(\vc{r}_n)\ \chi_{1/2 m_{s_n}}(n)\ \xi_{1/2 m_{t_n}}(n),\\
&&\phi_{\vc{k}_n}(\vc{r}_n) = \frac{1}{\sqrt{\Omega}} \exp ( i \vc{k}_n \cdot \vc{r}_n), 
\label{eq:spwf}
\end{eqnarray}
where $\gamma=(\vc{k},m_s,m_t)$ represents the quantum number of the single-nucleon wave function, and $\chi$ and $\xi$ are the spin and isospin wave functions, respectively.
The periodic boundary condition is imposed for the single-nucleon wave function in Eq.~(\ref{eq:spwf}), where the nucleon is confined in a box with length $L$ and volume $\Omega=L^3$. 
Then, the density is given as $\rho=2{k_F}^2/(3\pi^2)$

The correlation functions $F_S$ and $F_D$ in Eq.~(\ref{eq:correlation_fun}), which describe the spin-isospin dependent central correlation and tensor correlation in nuclear matter, respectively, are defined as
\begin{eqnarray}
F_S &=& \frac{1}{2} \sum_{i \not= j} f_S(i,j) = \frac{1}{2} \sum_{s=0}^{1}\sum_{t=0}^{1}\sum_{i \not= j} f_S^{(st)}(r_{ij})P^{(st)}_{ij}, 
\label{eq:fs}\\
F_D &=& \frac{1}{2} \sum_{i \not=j} f_D(i,j) = \frac{1}{2} \sum_{s=0}^{1}\sum_{t=0}^{1} \sum_{i \not= j}f_D^{(st)} (r_{ij}) r^2_{ij} S_{12}(i,j)P^{(st)}_{ij} \delta_{s1},
\label{eq:fd} 
\end{eqnarray}
where $S_{12}$ is the tensor operator, 
\begin{eqnarray}
&&S_{12}(i,j)= 3(\vc{\sigma}_i\cdot\hat{\vc{r}}_{ij})(\vc{\sigma}_j\cdot\hat{\vc{r}}_{ij}) - (\vc{\sigma}_i\cdot\vc{\sigma}_j)
\end{eqnarray}
with $\hat{\vc{r}}_{ij}=\vc{r}_{ij}/r_{ij}$ and $\vc{r}_{ij}=\vc{r}_{i} - \vc{r}_{j}$.
The operator $P^{(st)}_{ij}$ denotes the projection operator of the spin $s$ and isospin $t$ states of the $ij$-nucleon pair.

The Hamiltonian of the nuclear matter is expressed,
\begin{eqnarray}
{\cal H} = \sum_{i} t_i + \frac{1}{2} \sum_{i\not = j} v_{ij} + \frac{1}{6} \sum_{i\not= j \not=k} V_{ijk},
\end{eqnarray}
where $t$, $v$, and $V$ denotes, respectively, the kinetic energy, two-nucleon interaction, and three-nucleon interaction.

According to the cluster expansion theory under the present TOFS method discussed in Ref.~\cite{yamada19_cluster}, the binding energy per particle in nuclear matter, $B_{\rm ex}$, with use of the exponential type correlated wave function $\Psi_{\rm ex}$ in Eq.~(\ref{eq:correlation_fun}) is presented as 
\begin{eqnarray}
&&-B_{\rm ex} = \frac{1}{A}\, \frac{\langle \Psi_{\rm ex} | \mathcal{H} | \Psi_{\rm ex} \rangle }{\langle \Psi_{\rm ex} | \Psi_{\rm ex} \rangle } = \frac{1}{A}\, \frac{\left\langle \Phi_0 | \exp(F^{\dagger}) \mathcal{H} \exp(F) | \Phi_0 \right\rangle}{\left\langle \Phi_0 | \exp(F^{\dagger}) \exp(F) | \Phi_0 \right\rangle}
= \frac{1}{A}\, \sum_{n=0}^{\infty}\, (E_n)_{\rm c},
\label{eq:B_ex}
\\
&&(E_n)_{\rm c} = \sum_{\substack{n_1,n_2\\{n_1+n_2=n}}} \frac{1}{{n_1}!\ {n_2}!}\ {\left\langle \Phi_0 | F^{n_1}\mathcal{H}F^{n_2} | \Phi_0 \right\rangle}_{\rm c},
\label{eq:Enc}
\end{eqnarray}
where ${\left\langle \Phi_0 | F^{n_1}\mathcal{H}F^{n_2} | \Phi_0 \right\rangle}_{\rm c}$ is the summation of the linked diagrams in the matrix element of ${\left\langle \Phi_0 | F^{n_1}\mathcal{H}F^{n_2} | \Phi_0 \right\rangle}$ (see Ref.~\cite{yamada19_cluster}).
The cluster expansion theorem~\cite{yamada19_cluster} guarantees that the unlinked diagrams in each matrix element are completely canceled out in each order in the cluster expansion, and then only linked diagrams {remain} as shown in Eq.~(\ref{eq:Enc}). 
The expression (\ref{eq:B_ex}) means that the binding energy per particle, $B_{\rm ex}$, is expressed as the sum of only the linked diagrams.
Since each integral of the linked diagram is proportional to the nucleon number $A$ in nuclear matter, $B_{\rm ex}$ becomes $A$-independent or $\rho$-dependent~\cite{yamada19_cluster}.
\tred{
It is noted that the exponential expansion in Eqs.~(\ref{eq:B_ex}) and (\ref{eq:Enc}) is possible even when there are non-commutative operators such as tensor ones.
The proof is presented in Ref.~\cite{yamada19_cluster}.}

It is useful to discuss briefly the difference between the present TOFS theory and the coupled-\tred{cluster} (CC) theory. In the CC framework~\cite{baardsen13,hagen14,lietz17,kummel78,barlett81}, which is {\it non-Hermitian}, the energy per particle in nuclear matter is evaluated by $-B_{\rm CC} = \frac{1}{A} \langle \Phi_{0} | \exp(-\hat{S}) {\cal H} \exp(\hat{S}) | \Phi_0 \rangle$, where the cluster operator $\hat{S}$ is defined as the sum of $m$-particle $m$-hole excitation operators. The TOFS framework, however, is {\it Hermitian}, as shown in Eq.~(\ref{eq:B_ex}). 

In the present TOFS method, the radial parts of the correlation functions in Eqs.~(\ref{eq:fs}) and ($\ref{eq:fd}$) are expanded in terms of the Gaussian functions,
\begin{eqnarray}
&&f_S^{(st)}(r) = \sum_{\mu}C^{(st)}_{S,\mu} \exp\left[-a^{(st)}_{S,\mu} r^2\right],
\label{eq:exp_gs_FS}\\
&&f_D^{(st)}(r) = \sum_{\mu}C^{(st)}_{D,\mu} \exp\left[-a^{(st)}_{D,\mu} r^2\right].
\label{eq:exp_gs_FD}
\end{eqnarray}
Here, $C^{(st)}_{S,\mu}$ and $a^{(st)}_{S,\mu}$ together with $C^{(st)}_{D,\mu}$ and $a^{(st)}_{D,\mu}$ are the variational parameters.
They are determined so as to minimize the energy per particle in nuclear matter ($-B_{\rm ex}$) in Eq.~(\ref{eq:B_ex}).
It is noted that the Gaussian correlation functions bring about simplification and numerical stabilization for evaluating the matrix elements of many-body operators in the present nuclear matter calculation. 
The values of the size parameters, $a^{(st)}_{S,\mu}$ and $a^{(st)}_{D,\mu}$, are appropriately chosen by considering the range of the nuclear force and $k_F$, and only the expansion coefficients, $\left\{C_{S,\mu}^{(st)}\right\}$ and $\left\{C_{D,\mu}^{(st)}\right\}$, are taken as the variational parameters.
They are determined from the following conditions:
\begin{eqnarray}
\frac{\partial B_{\rm ex}}{\partial C^{(st)}_{S,\mu}}=0,\ \ \ \ \ 
\frac{\partial B_{\rm ex}}{\partial C^{(st)}_{D,\mu}}=0. 
\label{eq:system_eq}
\end{eqnarray}
The solutions to the system of equations $(\ref{eq:system_eq})$ for $C^{(st)}_{S,\mu}$ and $C^{(st)}_{D,\mu}$ give the variational minimization of the energy per particle in nuclear matter, $-B_{\rm ex}$, in Eq.~(\ref{eq:B_ex})\tred{, and determine the the radial correlation functions, $f^{(st)}_{S}(r)$ and $f^{(st)}_{D}(r)$, in Eqs.~(\ref{eq:fs}) and (\ref{eq:fd}), respectively.
Then, we get the (unnormalized) correlated nuclear matter wave function $\Psi_{\rm ex}$ in Eq.~(\ref{eq:correlation_fun}).
Since the wave function has an arbitrary constant factor, we can normalize the correlated nuclear matter wave function by using the constant factor.}
\tred{
It is noted that no supplementary conditions on the radial correlation functions $f^{(st)}_{S}(r)$ and $f^{(st)}_{D}(r)$ are required besides the conditions (\ref{eq:system_eq}) in the TOFS approach.
}

\tred{
Here, it is instructive to mention that the variational calculations with the Gaussian expansion method (GEM) have been widely used in few-body physics, hypernuclear physics, and nuclear cluster physics etc.~\cite{hiyama03,funaki08}, and precise calculations are possible for the energy and wave function of the few-body system etc. 
The GEM for the correlated functions presented in this paper is also employed in the finite nuclear calculations with the tensor-optimized antisymmetric molecular dynamics (TOAMD)~\cite{myo15,myo17_1}. 
TOAMD is a variational framework for the \textit{ab initio} description of light nuclei, where the correlation functions for the central-operator and tensor-operator types, $F_S$ and $F_D$, respectively, similar to eqs.~(\ref{eq:fs}) and (\ref{eq:fd}), are introduced and employed in a power-series form of the wave function. 
In the analysis of \textit{s}-shell nuclei with TOAMD, they have nicely reproduced the results of Green$^{\prime}$s function Monte Carlo (GFMC). 
It is noted that one can easily find out the minimum of the enegy of nuclear system by changing systematically the variational parameters in a wide model (or parameter) space. 
No additional conditions such as second derivatives are required in the variational calculations based on GEM, besides the conditions (\ref{eq:system_eq}).
}

In the numerical calculation \tred{of TOFS}, we approximate the exponential type correlated nuclear matter wave function $\Phi_{\rm ex}$ as the following $N$th-order power series type wave function $\Psi_N$,
\begin{eqnarray}
\Psi_N = \left[ \sum_{k=0}^{N} \frac{1}{k!} F^{k} \right] \Phi_0.
\label{eq:correlated_wf_finite}
\end{eqnarray}
It is noted that $\Psi_N$ at the limit of $N \rightarrow \infty$ becomes $\Psi_{\rm ex}$ in Eq.~(\ref{eq:correlation_fun}). 
Then, the binding energy per particle for $\Psi_N$ in Eq.~(\ref{eq:correlated_wf_finite}) is given as an approximation for $B_{\rm ex}$ in Eq.~(\ref{eq:B_ex}),
\begin{eqnarray}
&&{-B}_{N}
= \frac{1}{A}\,\frac{\left\langle \Psi_{N} \left| \mathcal{H} \right| \Psi_{N} \right\rangle}{\left\langle \Psi_{N} | \Psi_{N} \right\rangle} 
\simeq
\frac{1}{A}\,\sum_{n_1=0}^{N} \sum_{n_2=0}^{N} \frac{1}{{n_1}!\ {n_2}!}\ {\left\langle \Phi_{0} \left| F^{n_1}\mathcal{H}F^{n_2} \right| \Phi_{0} \right\rangle}_{\rm c},
\label{eq:BE_linked}
\end{eqnarray}
where only the linked diagrams in the matrix element are evaluated~\cite{yamada19_cluster}.
The expansion coefficients, $\left\{C_{S,\mu}^{(st)}\right\}$ and $\left\{C_{D,\mu}^{(st)}\right\}$, in Eqs.~(\ref{eq:exp_gs_FS}) and (\ref{eq:exp_gs_FD}), respectively, are taken as the variational parameters with respect to the energy $B_N$ in Eq.~(\ref{eq:BE_linked}).
They are determined from the following conditions:
\begin{eqnarray}
\frac{\partial B_{N}}{\partial C^{(st)}_{S,\mu}}=0,\ \ \ \ \ 
\frac{\partial B_{N}}{\partial C^{(st)}_{D,\mu}}=0. 
\label{eq:system_eq_N}
\end{eqnarray}
We refer to evaluating $B_{N}$ in Eq.~(\ref{eq:BE_linked}) as the $N$th-order TOFS calculation~\cite{yamada19_cluster}.


The TOFS formulation for pure neutron matter is similar to that for symmetric nuclear matter discussed above. 
Therefore, we skip the formulation in the present paper.

\section{Results and discussion}\label{sec3}

In the present paper, we perform the first-order TOFS calculation $(N=1)$ for symmetric nuclear matter and pure neutron matter up to the density of $\rho=0.5$ fm$^{-3}$ using the AV4' {\it NN} potential (central-type force) with short-range repulsion, where the correlated nuclear matter wave function $\Psi_{N=1}$ in Eq.~(\ref{eq:correlated_wf_finite}) is adopted, and the energy variational calculation is done with respect to $\Psi_{N=1}$ (see Eq.~(\ref{eq:system_eq_N})). 
It is noted that only the central-type correlation function $F_S$ in Eq.~(\ref{eq:fs}) is active in the present study, because the AV4' force is central.

In the present calculation, we study the contribution of the many-body terms from one-body to six-body ones which arise from the operator products~\cite{yamada19_cluster}, $F_S\mathcal{H} + \mathcal{H}F_S + F_S\mathcal{H}F_S$.
The energy per particle for symmetric nuclear matter (pure neutron matter), $E/A = -B_{N=1}$, in Eq.~(\ref{eq:BE_linked}), can be decomposed into the many-body terms,
\begin{eqnarray}
-B_{N=1} &=& \frac{E}{A} 
=\frac{1}{A}{\langle \Phi_0 | (1+F_S) \mathcal{H} (1+F_S) | \Phi_0 \rangle}_{\rm c} \nonumber \\
&=&e_{0}+{\sum_{m=2}^{6}e_{\rm mb}},
\label{eq:many_body_terms}\\
e_{0} &=& \frac{1}{A}\langle \Phi_{0} | \mathcal{H} | \Phi_{0} \rangle_{\rm c}, 
\label{eq:many_body_terms_e0}\\
e_{\rm mb} &=& {\left[\frac{1}{A}\langle \Phi_0 | F_S\mathcal{H} + \mathcal{H}F_S + F_S\mathcal{H}F_S | \Phi_0 \rangle_{\rm c}\right]}_{\rm mb}, 
\label{eq:many_body_terms_e2-e6}
\end{eqnarray} 
where $e_{0}$ denotes the energy per particle for the noncorrelated wave function $\Phi_{0}$ (Fermi-gas) of symmetric nuclear matter (pure neutron matter) , and $e_{\rm mb}$ stands for the contributions from the $m$-body terms \tred{($m=2, 3, 4, 5, 6$)} in the matrix element of $\frac{1}{A}\langle \Phi_0 | F\mathcal{H} + \mathcal{H}F + F\mathcal{H}F | \Phi_0 \rangle_{\rm c}$.

\begin{figure}[htbp]
\centering
\includegraphics[width=0.7\hsize]{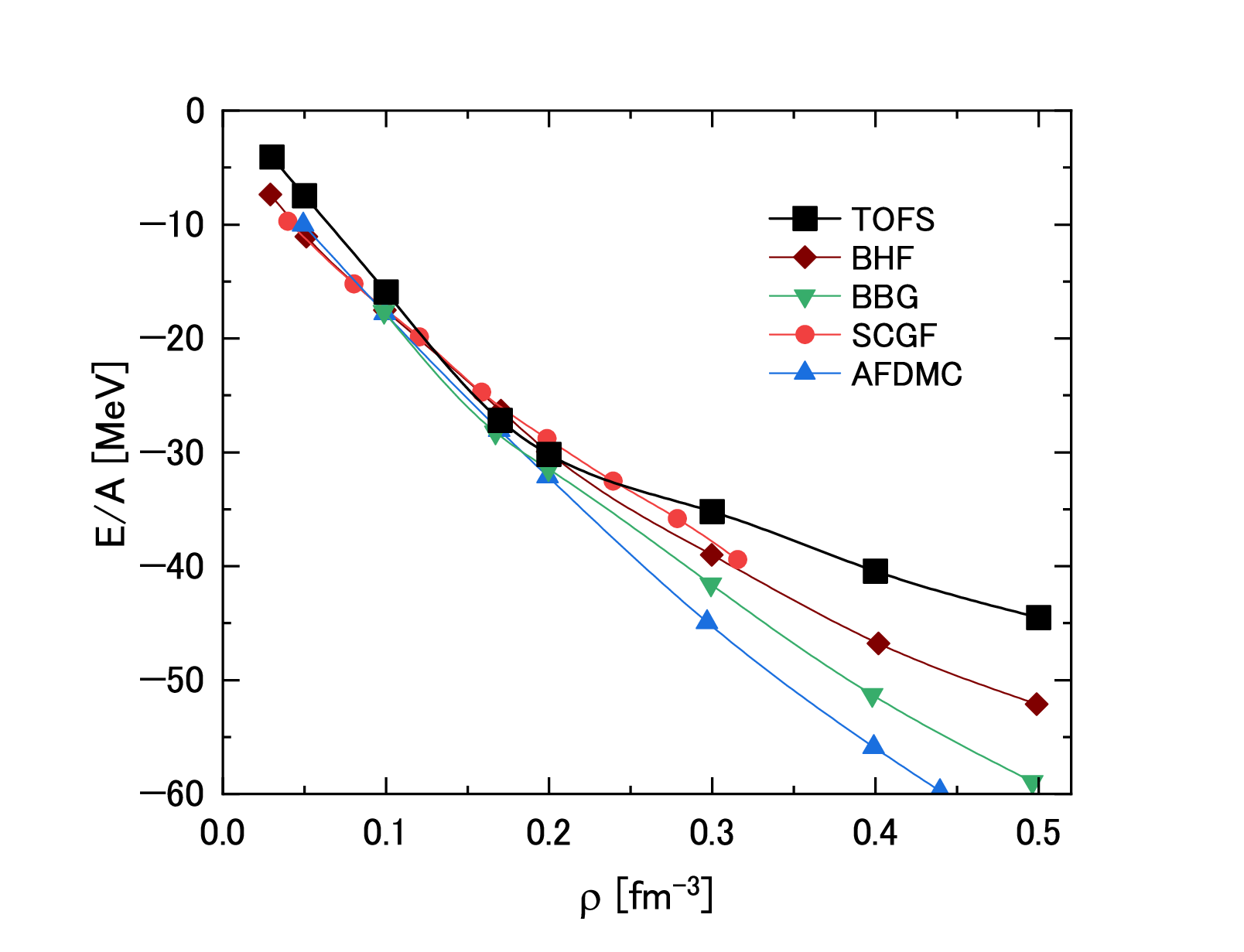}
\caption{
EOS of the TOFS method using the AV4' force for symmetric nuclear matter, compared with those of BHF, BBG, SCGF, and AFDMG~\cite{baldo12}.
}\label{fig:1}
\end{figure}

\subsection{Symmetric nuclear matter}

Figure~\ref{fig:1} shows the EOS of the TOFS method using the AV4' force for symmetric nuclear matter, compared with those of the other many-body methods (BHF, BBG, SCGF, \tred{AFDMC}).
In the density region less than $\rho \simeq 0.2$ fm$^{-3}$, the EOS of the TOFS calculation is similar to the other ones, while in the higher density region, it is a little stiffer than them.
It is noted that the EOS's for the BHF, BBG, SCGF, and \tred{AFDMC} methods also gradually deviate from each other beyond $\rho \simeq 0.2$~fm$^{-3}$. 
As mentioned in the previous sections, the TOFS theory is based on the variational method with respect to the correlated nuclear matter wave function $\Psi_{N=1}$ in Eq.~(\ref{eq:correlated_wf_finite}).
Therefore, the present results are considered to give the upper limit of EOS.

\begin{figure}[htbp]
\centering
\includegraphics[width=0.7\hsize]{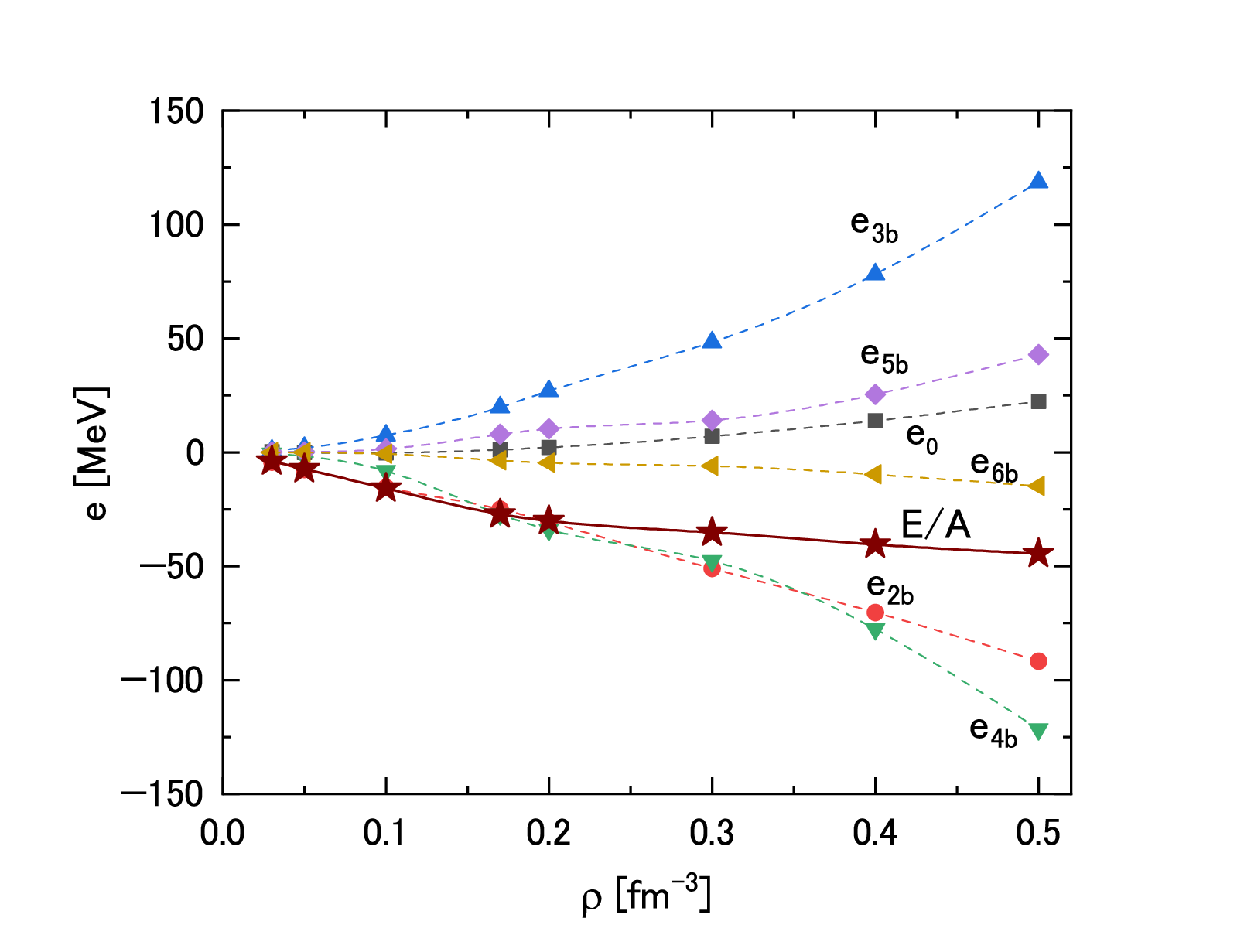}
\caption{
Density-dependent contributions from the many-body terms to $E/A$ for symmetric nuclear matter. 
See Eq.~(\ref{eq:many_body_terms_e2-e6}).
}\label{fig:2}
\end{figure}

It is interesting to see the density-dependent contributions from the many-body terms to the energy per particle for the symmetric nuclear matter ($E/A$) given in Eq.~(\ref{eq:many_body_terms_e2-e6}). 
The results are demonstrated in Fig.~\ref{fig:2}.
In the density region less than $\rho \simeq 0.1$ fm$^{-3}$, the many-body terms are small but non-negligible. However, increasing density $\rho$, their contributions to $E/A$ becomes gradually larger: for example, at $\rho=0.4$ fm$^{-3}$, $e_0 = 13.9$~MeV, $e_{\rm 2b} = -70.4$~MeV, $e_{\rm 3b} = 78.2$~MeV, $e_{\rm 4b} = -77.7$~MeV, $e_{\rm 5b} = 25.3$~MeV, $e_{\rm 6b} = -9.8$~MeV.
The contributions of $e_{\rm 2b}$, $e_{\rm 4b}$, and $e_{\rm 6b}$ are attractive, while those of $e_{0}$, $e_{\rm 3b}$, and $e_{\rm 5b}$ are repulsive.
These results indicate that one can not neglect even the $5$- and $6$-body terms in nuclear matter calculations.
The contributions from the \tred{higher-order} many-body terms may give a substantial effect on the momentum distribution of nucleon for nuclear matter. 

\begin{figure}[htbp]
\centering
\includegraphics[width=0.7\hsize]{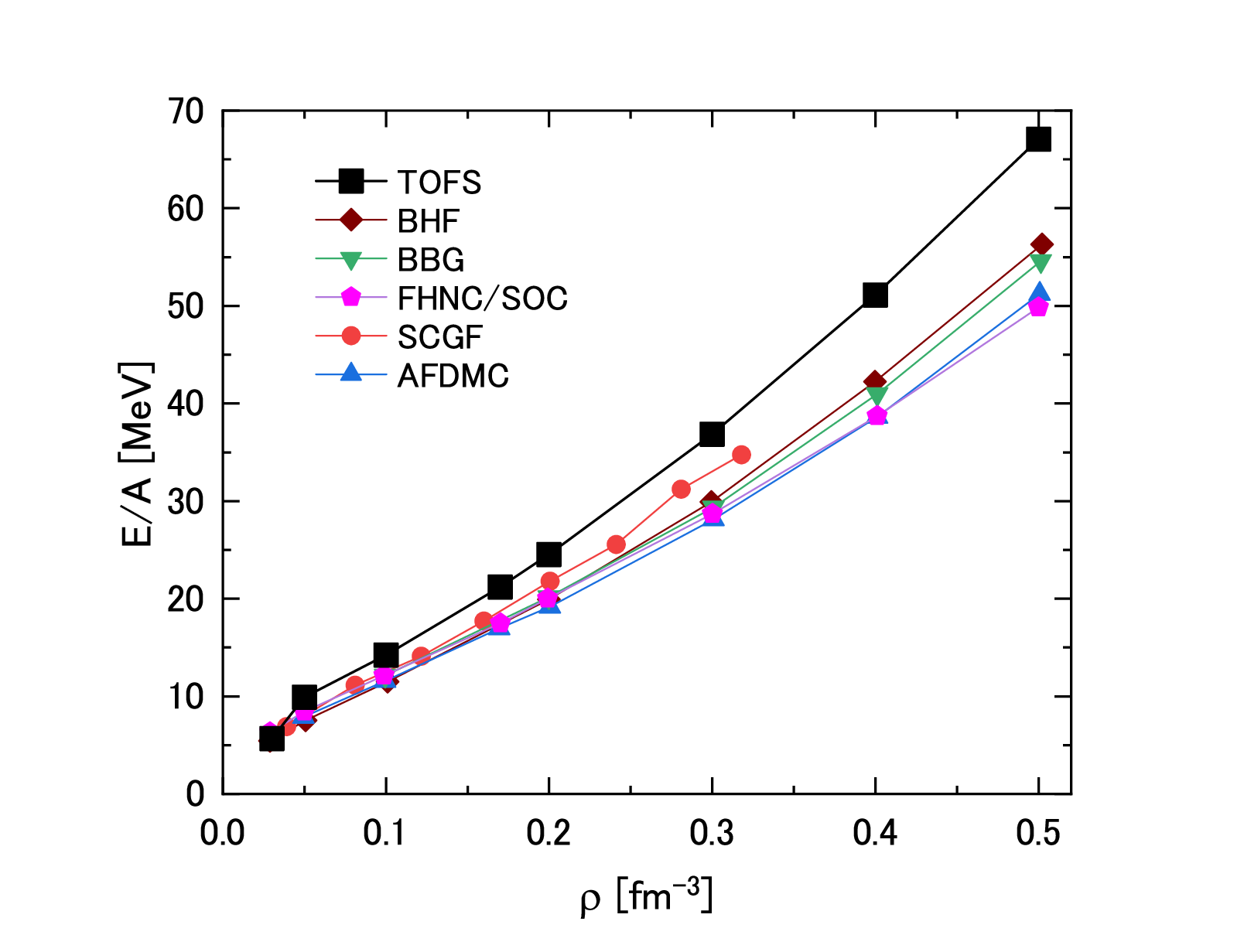}
\caption{
EOS of the TOFS method using the AV4' force for pure neutron matter, compared with those of BHF, BBG, FHNC/SOC, SCGF, and \tred{AFDMC}~\cite{baldo12}.
}\label{fig:3}
\end{figure}

\subsection{Pure neutron matter}

In pure neutron matter, only the isospin $T=1$ section of the {\it NN} force is active.
The EOS of the TOFS method using the AV4' force for pure neutron matter is displayed in Fig.~\ref{fig:3}, together with the other many-body methods (BHF, BBG, FHNC/SOC, SCGF, AFDMC).
Although the TOFS result is almost the same as the other ones at $\rho=0.03$~fm$^{-3}$, the TOFS's EOS is \tred{likely to be} a bit stiffer than the other methods.
This is related to the fact that the TOFS method gives an upper limit, because the method is variational with respect to the correlated pure neutron matter wave function.
\tred{However, the higher-order TOFS calculation might give some energy gain, the study of which is one of our future subjects.}
One can see that the EOS's for the BHF, BBG, FHNC/SOC, SCGF, and \tred{AFDMC} methods don't fully converge each other, and their numerical discrepancies widen gradually, when the density increases from \tred{$\rho=0.3$}~fm$^{-3}$.

\begin{figure}[htbp]
\centering
\includegraphics[width=0.7\hsize]{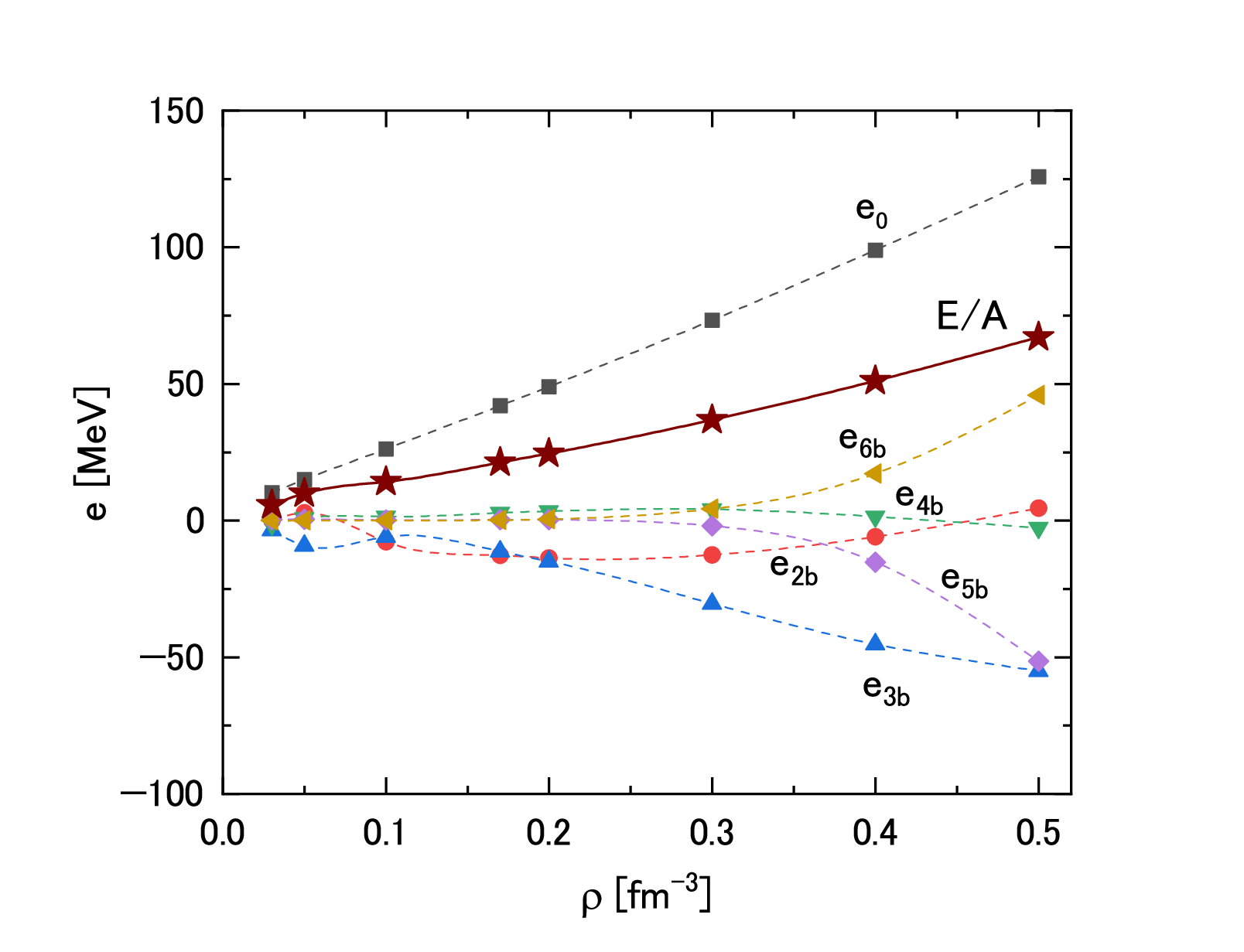}
\caption{
Density-dependent contributions from many-body terms to $E/A$ for pure neutron matter. 
See Eq.~(\ref{eq:many_body_terms_e2-e6}).
}\label{fig:4}
\end{figure}

The density-dependent contributions from the many-body terms to the energy per particle for pure neutron matter ($E/A$) are shown in Fig.~\ref{fig:4}.
In the low density region less then $\rho \simeq 0.3$~ fm$^{-3}$, the contributions to $E/A$ come dominantly from $e_{0}$, $e_{\rm 2b}$ and $e_{\rm 3b}$. However, beyond $\rho \simeq 0.3$~ fm$^{-3}$, those from $e_{\rm 5b}$ and $e_{\rm 6b}$ become gradually larger and can not be neglected anymore beyond $\rho \simeq 0.4$~fm$^{-3}$, though that from $e_{\rm 4b}$ is smaller than those from $e_{\rm 5b}$ and $e_{\rm 6b}$.

\tred{
Here, it is instructive to discuss how many-body terms should be considered to ensure the convergence in $E/A$ for both symmetric and pure neutron matters.
In the first-order TOFS calculation, there are the many-body terms up to 6-body one, as shown in Eq.~(\ref{eq:many_body_terms}).
The 6-body term is originated from the operaror product of $F_S V F_S$ (2nd power of $F_S$), where $V$ denotes the two-body \textit{NN} interactions.
In the second-order TOFS calculation ($N=2$ in Eq.~(\ref{eq:BE_linked})), one gets the many-body terms up to 10-body one, arising from the operator product of $F_S^2 V F_S^2$ (4th power of $F_S$)~\cite{yamada19_cluster}.
In general, it is expected that the contribution of the higher-order many-body terms than the 6-body one will become smaller. 
The reasons are given as follows:\ 
The higher-order many-body terms beyond 6-body one come from the product of $V$ and the 3rd (4th, and higher-order) power of $F_S$.
The effect of the correlation function $F_S$ is expected to be smaller than 1. 
Therefore, the contribution of the 3rd and 4th etc. powers of $F_S$ is expected to be smaller than that of the 2nd power of $F_S$. 
This fact suggests that the many-body terms obtained in the first order TOFS calculation are considered to give significant contributions, although the higher order TOFS calculations may provide more enough convergence.
This should be confirmed by numerical calculations in future.
}

\section{Summary}\label{sec4}
We have studied the EOS of the symmetric nuclear matter and pure neutron matter up to $\rho=0.5$~fm$^{-3}$ with the TOFS method, being the first principle theory.
This is the variational method with respect to the correlated nuclear matter wave function, which is described by the exponential type and, its approximation, the power-series type, satisfying the linked-cluster expansion theorem.
The spin-isospin dependent central AV4' {\it NN} potential with the short-range repulsion was used, for the purpose of the benchmark calculations.
The EOS's obtained by the first-order TOFS method were reasonably reproduced, compared with the other \textit{ab initio} many-body methods.
The present TOFS results are considered to give the upper limits for the EOS's, because the TOFS method is variational. 
In higher density the many-body terms are found to give sizable effects on $E/A$ and play an important role in nuclear matter.
The non-negligible contributions from the many-body terms may give a substantial effect on the momentum distribution of nucleon in nuclear matter. 

The TOFS method is interesting and promising to study the nuclear matter with the first principle theory.
This method is useful to clarify the role of the many-body correlations in nuclear matter, originating from the short-range repulsion, tensor component etc.\ of the \textit{NN} interaction, as well as the antisymetrization among nucleons.
In this sense, it is important to perform the TOFS calculations with use of the AV'6, AV'8 and AV18 forces, including the three-nucleon force.

\section*{Acknowledgments}

This work was partially supported by the JSPS KAKENHI Grant Numbers JP26400283, JP23K03397.


\end{document}